\documentclass[copyright,creativecommons]{eptcs}
\providecommand{\event}{GaM 2015} 

\usepackage{amsmath}
\usepackage{amssymb}
\usepackage{amsthm}
\usepackage{multirow}
\usepackage{array}
\usepackage{hyperref}
\usepackage{caption}
\usepackage{subcaption}
\usepackage{stmaryrd}
\usepackage{cleveref}
\usepackage{enumerate}
\usepackage{MnSymbol}
\usepackage{graphicx}
\usepackage[all]{xy}
\usepackage{wrapfig}

\newcommand{\Nat}{\mathbb{N}}
\newcommand{\Vals}{\mathbb{D}}
\newcommand{\RealInf}{\mathbb{R}_{\infty}}
\newcommand{\Parts}{\mathcal{P}}

\newcommand{\names}{\ensuremath{\mathcal{N}}}
\newcommand{\ealph}{\ensuremath{\mathcal{E}}}
\newcommand{\tequiv}{\equiv}
\newcommand{\deng}[1]{\left\llbracket #1 \right\rrbracket^g}
\newcommand{\denc}[1]{\left\llbracket #1 \right\rrbracket^c}
\newcommand{\cost}[1]{\llangle #1 \rrangle}
\newcommand{\funres}[2]{{#1}_{|#2}}
\newcommand{\ngraph}[2]{{#1}\rhd {#2}}
\newcommand{\nhg}{NH-graph}

\newcommand{\axpar}{{\bf (\textsc{AX}$_\parallel$)}}
\newcommand{\axres}{{\bf (\textsc{AX}$_{(x)}$)}}
\newcommand{\axalfa}{{\bf (\textsc{AX}$_\alpha$)}}
\newcommand{\axse}{{\bf (\textsc{AX}$_{SE}$)}}
\newcommand{\axperm}{{\bf (\textsc{AX}$_\pi$)}}
\newcommand{\axpermdis}{{\bf (\textsc{AX}$_\pi^p$)}}

\newtheorem{theorem}{Theorem}
\newtheorem{lemma}{Lemma}
\newtheorem{remark}{Remark}
\newtheorem{proposition}{Proposition}
\newtheorem{definition}{Definition}
\newtheorem{property}{Property}

\crefname{property}{property}{properties}

\title{Dynamic Programming on Nominal Graphs\thanks{Research supported by the EU Integrated Project 257414 ASCENS and by the Italian MIUR Project CINA (PRIN 2010), grant number 2010LHT4KM.}}
\author{
Nicklas Hoch 
\institute{Volkswagen AG, Corporate Research Group}
\email{nicklas.hoch@volkswagen.de}
\and
Ugo Montanari 
\institute{University of Pisa, Computer Science Department}
\email{ugo@di.unipi.it}
\and Matteo Sammartino
\institute{Radboud University, Institute for Computing and Information Sciences}
\email{m.sammartino@cs.ru.nl}
}
\def\titlerunning{Dynamic Programming on Nominal Graphs}
\def\authorrunning{N. Hoch \& U. Montanari \& M. Sammartino}
\begin{document}
\maketitle

\begin{abstract}
Many optimization problems can be naturally represented as (hyper) graphs, where vertices correspond to variables and edges to tasks, whose cost depends on the values of the adjacent variables. Capitalizing on the structure of the graph, suitable dynamic programming strategies can select certain orders of evaluation of the variables which guarantee to reach both an optimal solution and a minimal size of the tables computed in the optimization process.
In this paper we introduce a simple algebraic specification with parallel composition and restriction whose terms up to structural axioms are the graphs mentioned above. In addition, free (unrestricted) vertices are labelled with variables, and the specification includes operations of name permutation with finite support. We show a correspondence between the well-known tree decompositions of graphs and our terms. If an axiom of scope extension is dropped, several (\emph{hierarchical}) terms actually correspond to the same graph. A suitable graphical structure can be found, corresponding to every hierarchical term. Evaluating such a graphical structure in some target algebra yields a dynamic programming strategy. If the target algebra satisfies the scope extension axiom, then the result does not depend on the particular structure, but only on the original graph. 
We apply our approach to the parking optimization problem developed in the ASCENS e-mobility case study, in collaboration with Volkswagen. Dynamic programming evaluations are particularly interesting for autonomic systems, where actual behavior often consists of propagating local knowledge to obtain global knowledge and getting it back for local decisions. 
\end{abstract}

\section{Introduction}

Many optimization problems can naturally be represented as hypergraphs, where each hyperedge is an atomic subproblem and is attached to vertices corresponding to the problem's variables. However, these hypergraphs often lack an algebraic structure, which would allow for the recursive resolution of problems, and are not able to represent the \emph{secondary optimization problem}, that is: finding an optimal \emph{variable elimination strategy}. The order in which variables are eliminated may dramatically affect the computation cost of solutions.

In this paper we introduce a simple algebraic specification for representing optimization problems. It is similar to a process calculus, based on nominal structures (namely \emph{permutation algebras}, see e.g.\ \cite{nomsns,GadducciMM06}). Optimization problems are represented as terms over the variables of the problem, consisting of the parallel composition of subproblems. A key feature is the representation of variable elimination via the restriction operator. In fact, since a restricted variable can only occur in its scope, its value can be determined when solving the subproblem it encloses. For instance
\[
	(x_1) ( (x_2) A(x_1,x_2) \parallel (x_3) B(x_1,x_3) )
\]
represents a problem with three variables $x_1,x_2,x_3$ and two atomic problems $A,B$, sharing $x_1$. One can solve the subproblem $(x_2) A(x_1,x_2)$ w.r.t.\ $x_2$ right away, thus eliminating $x_2$ and obtaining a solution $A'(x_1)$ parametric w.r.t.\ $x_1$; similarly for $(x_3) B(x_1,x_3)$. Then $x_1$ can be eliminated in $(x_1) ( A'(x_1) \parallel B'(x_1) )$, which yields the global solution.

The specification provides axioms that are common in process algebras, and have a natural interpretation in terms of optimization problems. For instance: subproblems can be solved in any order, the formal names of their variables are irrelevant and so on. In particular, \emph{scope extension} (actually, scope reduction) becomes a crucial operation: it allows variables to be eliminated \emph{earlier}, when solving a smaller subproblem. This produces a more efficient solution of the secondary optimization problem of dynamic programming.

The computation of an optimal solution is formalized as an evaluation of terms in a suitable domain of \emph{cost functions}, giving a cost to each assignment of free variables. This domain is indeed an algebra for the specification: constructors are interpreted as operations over cost functions, performing optimization steps, and axioms become useful properties. In particular, they tell that cost functions are preserved under rearranging the structure of the problem to get a more efficient computation. Moreover, the underlying nominal structure provides a notion of \emph{support}, that is the set of variables that are really relevant for the computation. This is the key for a finite, efficient representation of cost functions.

Then we introduce a graphical notation for optimization problems, called \emph{nominal hypergraphs}. They are hypergraphs with an \emph{interface} where names are assigned to some vertices. These vertices are variables for the overall problem. Other vertices are variables of subproblems that must be eliminated in the optimization process.

We show that nominal hypergraphs can be given an algebraic structure, and that (isomorphic) nominal hypergraphs can be described by (congruent) terms of our specification. This allows us to recursively compute the cost function of a nominal hypergraph by performing the computation on any of the corresponding terms. Moreover, we show that well-known structures to represent parsing of graphs, namely \emph{tree decompositions} \cite{RobertsonS84,Kloks94}, can be represented in our framework as terms.

However, nominal graphs still lack a description of the variable elimination strategy. We describe this information as \emph{hierarchical} nominal hypergraphs, that are trees describing the decomposition of a nominal hypergraph in terms of nested components, each corresponding to a subproblem. Such trees correspond to terms without the scope extension axioms, i.e., where the scope of restrictions is fixed. A bottom-up visit of a hierarchical nominal graph yields a dynamic programming algorithm with the given variable elimination strategy.

We apply our approach to the \emph{e-mobility case study} from the \emph{Autonomic Service-Component Ensembles} (ASCENS) project. In ASCENS, systems are modeled as self-aware, self-adaptive and autonomic 
components running in ensembles (dynamic aggregations of components), which through interactions among them and with the environment accomplish both individual (local)
and collective (global) goals by optimizing the use of resources.

In the e-mobility case study, carried on in collaboration with Volkswagen, the traffic system is modeled as ensembles of electrical vehicles with the goal to optimize the usage of resources (electricity, parking places, etc.), while ensuring the fulfillment of individual goals (such as reaching in time the destination) and collective goals (avoiding traffic jams or guaranteeing that all vehicles find a spot where to park).
So in general, besides optimizing local resources, for example by finding the best trips and journeys for each vehicle, the e-mobility case study aims at solving global problems, involving large ensembles of different vehicles. In \cite{HochZWS12} several optimization problems are presented for the e-mobility case study. In \cite{BuresNGHKKMMPSWZ13} a parking problem is considered. In the formulation we present in this paper, parking systems are regarded as nominal hypergraphs, where each hyperedge is a parking zone and vertices are cars that may be parked inside the zone. Their term representation is interpreted as functions telling the cost of parking cars inside or outside certain parking zones. Thus cost functions only have binary arguments. While more efficient, this domain choice yields operators remarkably neater than those of the classical point-wise interpretation and our framework is flexible enough to accommodate them.

\section{An algebraic specification for optimization problems}
\label{sec:desc}

We introduce an algebraic specification for describing the structure of optimization problems. Variables are represented as \emph{names}, belonging to an enumerable set \names. We write $Perm(\names)$ for the set of permutations over \names, i.e., bijective functions $\pi \colon \names \to \names$.
\begin{definition}[Optimization signature]
Let $C$ be a set of constants denoting atomic problems, equipped with an arity function $ar: C \to \Nat$ telling how many variables each problem involves. We assume the \emph{empty problem} $nil$, with $ar(nil) = 0$. The \emph{optimization signature} is given by the following grammar
\[
  p,q:= p \parallel q\ \mid \ (x)p \ \mid \ p \pi \ \mid \ A(\tilde{x}) \ \mid \ nil   
\]
where $A \in C$, $\pi \in Perm(\names)$, $\{x\} \cup \tilde{x} \subseteq \names$ and $|\tilde{x}| = ar(A)$ (we overload the notation $\tilde{x}$ to indicate both a vector and a set of names). 
\label{def:ng-sig}
\end{definition}
Here: 
\begin{itemize}
    \item the \emph{parallel composition} $p \parallel q$ represents the problem consisting of two subproblems $p$ and $q$, possibly sharing some variables;
    \item the \emph{restriction} $(x)p$ is $p$ where the assignment for $x$ has already been determined;
    \item the \emph{permutation} $p \pi$ is $p$ where variable names have been exchanged according to $\pi$;
    \item the \emph{atomic problem} $A(\tilde{x})$ represents a problem that only involves the problem $A$ over variables $\tilde{x}$; 
    \item $nil$ represents the \emph{empty problem}.
\end{itemize}
We assume restriction has precedence over parallel composition.

Free names of $p$ are recursively defined as follows
\begin{align*}
    fn(p \parallel q) &= fn(p) \cup fn(q) &
    fn((x)p) &= fn(p) \setminus \{x\} &
    fn(p \pi) &= \pi(fn(p)) \\
    fn(A(\tilde{x})) &= \tilde{x} & fn(nil) &= \emptyset 
\end{align*}

We consider syntax up to structural congruence axioms shown in \cref{fig:terms-congr}.
\begin{figure}[t]
\begin{alignat*}{3}
	\text{\axpar} 
	& \qquad && p \parallel q \tequiv q \parallel p 
	\qquad  (p \parallel q) \parallel r \tequiv p \parallel (q \parallel r) 
	\qquad 
	p \parallel nil \tequiv p
	\\[1ex]
	\text{\axres} 
	& \qquad &&
	(x)(y) p \tequiv (y)(x)p 
    \qquad
    (x) nil \tequiv nil
	\\[1ex]
    \text{\axalfa} 
    & \qquad &&
	(x)p \tequiv (y)p[x \mapsto y] \qquad (y \notin fn(p))
	\\[1ex]
	\text{\axse}
	& \qquad &&
	(x) ( p \parallel q ) \tequiv (x) p \parallel q  \qquad (x \notin fn(q))
	\\[1ex]
	\text{\axperm}
	& \qquad &&
	p \; \mathsf{id} \tequiv p 
	\qquad 
	( p \pi') \pi \tequiv p ( \pi \circ \pi' )
	\\[1ex]
	\text{\axpermdis}
	& \qquad &&
    \begin{gathered}
A(x_1,\dots,x_n) \pi \tequiv A(\pi(x_1),\dots,\pi(x_n))
	\qquad
	nil \, \pi \tequiv nil 
	\qquad
	(p \parallel q)\pi \tequiv p \pi \parallel q \pi 
	\\
	( (x)p ) \pi \tequiv (x) p \pi' \qquad (\pi'(x) = x, \text{$\pi'(y) = \pi(y)$ for $x \neq y$})
	\end{gathered}
\end{alignat*}
\caption{Structural congruence axioms of the optimization specification.}
\label{fig:terms-congr}
\end{figure}
The operator $\parallel$ forms a commutative monoid, meaning that problems in parallel can be solved in any order \axpar. 
Restrictions can be $\alpha$-converted \axalfa, i.e.\ names of assigned variables are irrelevant. Restrictions can also be swapped, i.e., assignments can happen in any order, and removed, whenever their scope is $nil$ \axres. The scope of restricted variables can be narrowed to terms where they occur free \axse. Axioms regarding permutations say that identity and composition behave as expected \axperm\ and that permutations distribute over syntactic operators \axpermdis. Permutations are assumed to behave in a capture avoiding way when applied to $(x)p$. We call \emph{optimization algebraic specification} the specification made of the optimization signature and the congruence axioms, and \emph{optimization terms} the terms for the specification.

We include permutations in the specification because they provide a general mechanism to compute the set of ``free'' names in any algebra, called \emph{(minimal) support}.
\begin{definition}[Support]
Let $A$ be an algebra for the optimization specification, and let $\pi^A$ be the interpretation of $\pi$ in $A$. We say that $X \subset \names$ \emph{supports} $a \in A$ whenever, for all permutations $\pi$ acting as the identity on $X$, we have $a \pi^A = a$. The minimal support $supp(a)$ is the intersection of all sets supporting $a$.
\end{definition}
For instance, let $\pi^t$ be the interpretation of $\pi$ on optimization terms: given a term $p$, $p \pi^t$ applies $\pi$ to all free names of $p$ in a capture avoiding way. It is easy to verify that $supp(p) = fn(p)$.

\subsection{Hierarchical optimization specification}

The scope of restrictions determines a solution for the \emph{secondary optimization problem}, because it specifies when restricted variables should be eliminated. However, the presence of \axse\ identifies terms corresponding to different solutions. We call \emph{hierarchical optimization specification} the optimization specification without \axse, and hierarchical terms its freely generated terms.

We are interested in two forms of hierarchical terms.
\begin{definition}[Normal and canonical forms]
A term is said to be in \emph{normal form} whenever it is of the form
\[
    (\tilde{x})( A_1(\tilde{x}_1) \parallel A_2(\tilde{x}_2) \parallel \dots \parallel A_n(\tilde{x}_n) )
\]
with $A_i \in C$ ($i=1,\dots,n$) and $\tilde{x} \subseteq \tilde{x}_1 \cup \dots \cup \tilde{x}_n$.
It is in \emph{canonical form} whenever it is obtained by the repeated application to a non-hierarchical term of \axse, from left to right, until termination. For both forms, we assume that subterms of the form $(\tilde{x}) nil$ (where $\tilde{x}$ may be empty) are removed using \axres{} and \axpar.
\end{definition}
Normal and canonical forms are somewhat dual: normal forms have all restrictions at the top level, whereas in canonical forms every restriction $(x)$ is as close as possible to the atomic terms where $x$ occurs (if any). A term in normal form is intuitively closer to a typical optimization problem: $\tilde{x}$ specifies which variables should be assigned, and the term in its scope represents subproblems and their connections. In a term in canonical form, variables are eliminated as soon as possible. Notice that a term may have more than one canonical form, whereas normal forms are unique (up to the hierarchical optimization specification congruence).

\begin{remark}
\label{rem:can-rep}
Hierarchical terms in normal and canonical form can be regarded as canonical representatives of $\tequiv$-classes (recall that $\tequiv$ is the structural congruence of \cref{fig:terms-congr}), because $\tequiv$ is coarser than the hierarchical optimization specification congruence.
\end{remark}

\section{Optimization problems as nominal hypergraphs}
\label{sec:ngraphs}

Recall that a hypergraph $G$ is a triple $(V_G,E_G,a_G \colon E_G \to V^\star_G)$, where $V_G$ is the set of vertices, $E_G$ is the set of hyperedges and, for each $e \in E_G$, $a_G(e)$ is the tuple of vertices attached to $e$ ($V^\star_G$ is the set of tuples over $V_G$). Let \ealph\ be a set of edge labels, equipped with a function $ar \colon \ealph \to \Nat$ telling the number of vertices $ar(l)$ of an edge with label $l$. A \emph{labeled} hypergraph $G$ is a hypergraph $G$ plus a function $lab \colon E_G \to \ealph$ mapping each hyperedge $e \in E_G$ to its label $l$ such that $|a_G(e)| = arl(l)$. Given two (labeled) hypergraphs $G_1$ and $G_2$, we write $G_1 \uplus G_2$ for their component-wise disjoint union. 

Optimization problems can naturally be seen as hypergraphs labeled over atomic subproblems, where vertices correspond to variables. 
We introduce a notion of labeled hypergraph where some vertices are associated variable names.
\begin{definition}[Labeled nominal hypergraph and their morphisms]
A \emph{labeled nominal hypergraph} (\nhg\ in short) is a pair $\ngraph{\eta}{G}$, where $G$ is a labeled hypergraph without isolated vertices and $\eta$ is a partial injection from $V_G$ to $\names$, assigning names to some vertices of $G$. The set $img(\eta)$ is called the \emph{interface} of $G$ and $def(\eta)$ (the domain of definition of $\eta$) are called \emph{interface vertices}. Given two \nhg s $\ngraph{\eta_1}{G_1}$ and $\ngraph{\eta_2}{G_2}$, a \nhg\ morphism $h \colon \ngraph{\eta_1}{G_1} \to \ngraph{\eta_2}{G_2}$ is a homomorphism $G_1 \to G_2$ of labeled hypergraphs that preserves names, namely $\eta_1 \circ h_V = \eta_2$, where $h_V$ is the action of $h$ on vertices.
\end{definition}
Interface vertices can be understood as ``external'' vertices, with a public, global identity. They may be interaction points, i.e., they may be shared, with other graphs. This will allow for a simple definition of parallel composition of \nhg s. Notice that \nhg{} homomorphisms must be injective on vertices, because they must commute with functions that are injective on vertices.

We say that $\ngraph{\eta_1}{G_1}$ and $\ngraph{\eta_2}{G_2}$ are isomorphic, written $\ngraph{\eta_1}{G_1} \cong \ngraph{\eta_2}{G_2}$, whenever there is an \nhg\ isomorphism (i.e., a \nhg\ morphism whose underlying hypergraph homomorphism is an isomorphism) between them.

\begin{remark}
\label{rem:span}
A \nhg\ $\ngraph{\eta}{G}$ can be seen as the following span of (total) injective graph homomorphisms
\[
    \xymatrix{
        [\names] & [img(\eta)] \ar@{_{(}->}[l]_-{\eta_l} \ar@{^{(}->}[r]^-{\eta_r} & G
    }
\]
where $[img(\eta)]$ is the discrete graph with vertices $img(\eta)$, $\names$ is the infinite discrete graph with vertices $\names$, $\eta_l(v) = \eta(v)$ and $\eta_r$ is an embedding.
\end{remark}

\subsection{Example}
Consider the optimization term shown in the introduction, without the outer restriction
\[
    (x_2) A(x_1,x_2) \parallel (x_3) B(x_1,x_3)
\]
Recall that such a term may represent the following optimization problem: given two subproblems $A$ and $B$, with cost functions parametric in $x_1$, $x_2$, and $x_1$, $x_3$, respectively, find the optimal total cost. Actually, here $x_1$ is free, meaning that the total cost is parametric in $x_1$. 

The problem can be represented as a \nhg\ $\ngraph{\eta}{G}$ with two hyperedges, labeled $A$ and $B$, and three 
\begin{wrapfigure}{r}{.2\textwidth}
\centering
\includegraphics[trim=50pt 50pt 50pt 0pt,scale=.35]{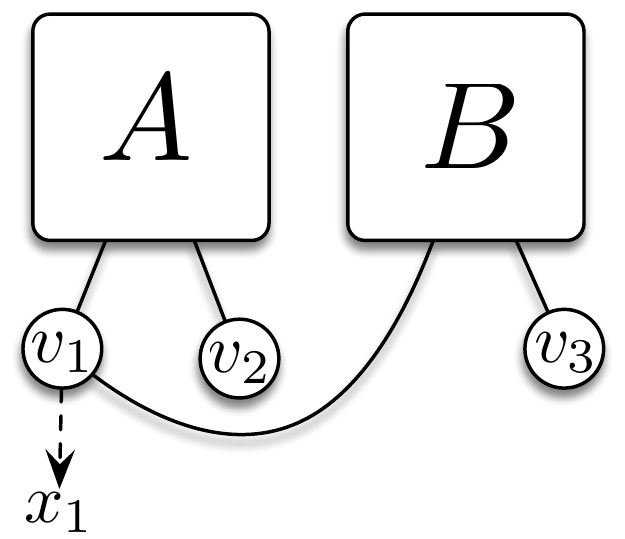}
\end{wrapfigure}
vertices $v_1,v_2,v_3$, corresponding to $x_1,x_2$ and $x_3$. 
Actually, only $x_1$ becomes an interface name, because it is the only variable the problem ``exposes''. 
Other variables are not part of the interface, meaning that they are taken up to $\alpha$-conversion. The \nhg\ is depicted on the right: the dashed line describes the domain of definition of $\eta$, namely $\eta(v_1) = x_1$.

\subsection{An algebra for \nhg s}

Now we show that we can regard \nhg s as elements of an algebra for the optimization specification. This will allow us to recursively evaluate (parsing of) \nhg s as cost functions, as done in \cref{sec:costfun-alg}.

Operations are interpreted as follows
\begin{align*}
    \ngraph{\eta_1}{G_1} \parallel^g \ngraph{\eta_2}{G_2} = \ngraph{\eta}{( G_1 \uplus G_2 )_{/\sim_V}} &
    \quad \text{where} \
    \left\{
    \begin{aligned}
        \sim_V & = \left\{ (v_1,v_2) \in V_{G_1} \times V_{G_2} \left|
        \begin{gathered}
            \eta_1(v_1) = \eta_2(v_2) \\
            \neq \text{undefined} 
        \end{gathered}
        \right.
        \right\}
        \\[1ex]
        \eta([v]_{\sim_v}) & = 
        \begin{cases}
            \eta_1(v) & v \in V_{G_1} \\
            \eta_2(v) & v \in V_{G_2}
        \end{cases}
    \end{aligned}
    \right.
    \\[2ex]
    (x)^g ( \ngraph{\eta}{G} ) = \ngraph{\eta_{\setminus x}}{G}&
    \quad
    \text{where}
    \quad
    \eta_{\setminus x}(v) =
    {
    \begin{cases}
        \text{undefined} & \eta(v) = x \\
        \eta(v) & \text{otherwise}
    \end{cases}
    }
    \\
    ( \ngraph{\eta}{G} ) \pi^g = \ngraph{(\pi \circ \eta)}{G}&
\end{align*}
The parallel composition $\ngraph{\eta_1}{G_1} \parallel_g \ngraph{\eta_2}{G_2}$ is computed by taking the disjoint union of 
the two \nhg s and then identifying vertices with the same interface names (formally, $/\sim_V$ takes equivalence classes of vertices). The function $\eta$ is defined on equivalence classes of vertices as expected. The restriction $(x)_g \ngraph{\eta}{G}$ of $\ngraph{\eta}{G}$ w.r.t.\ $x$ simply removes $x$ from the interface of $\ngraph{\eta}{G}$.

The interpretation of constants can be defined via a mapping
\newcommand{\AH}{
 {\mathchoice
  {\raisebox{-4.5ex}{\includegraphics[height=12ex]{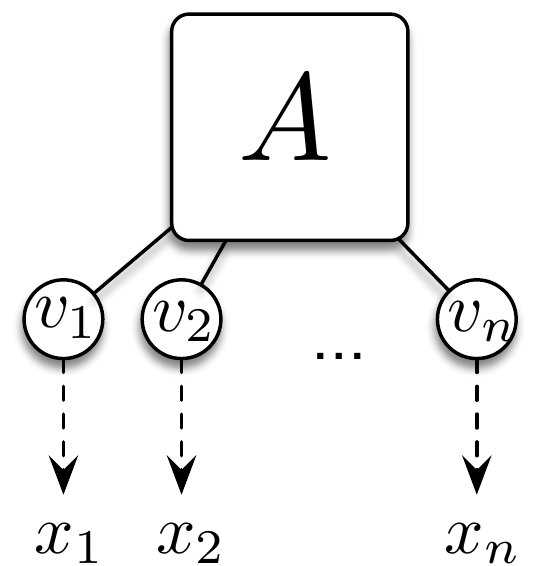}}}
  {\includegraphics[height=1.6ex]{images/H-const}}
  {\includegraphics[height=1.2ex]{images/H-const}}
  {\includegraphics[height=0.9ex]{images/H-const}}
 }
}
\[
    \deng{A(x_1,x_2,\dots,x_n)} = \AH
    \qquad\qquad\qquad
    \deng{nil} = \ngraph{! \colon \emptyset \to \names}{0_G}
\] 
where $! \colon \emptyset \to \names$ is the (unique) mapping from $\emptyset$ to $\names$ and $0_G$ is the empty hypergraph

Now we have to show that we have indeed defined an algebra. We have to check that congruence axioms are satisfied. We first need the following characterization of the minimal support of a \nhg.
\begin{lemma}
$supp(\ngraph{\eta}{G}) = img(\eta)$.
\label{lem:nhg-supp}
\end{lemma}
\begin{proposition}
Operations $\parallel^g$, $(x)^g$, $\pi^g$ satisfy axioms of \cref{fig:terms-congr}, where $\tequiv$ becomes $\cong$ and $fn(-)$ becomes $supp(-)$.
\label{prop:nhg-ops-axioms}
\end{proposition}
Now we can define a unique evaluation of optimization terms: given $p$, the corresponding \nhg\ $\deng{p}$ can be computed by structural recursion using the evaluation of constants given above and the interpretation of operations on \nhg s. This induces a sound and complete axiomatization for \nhg s. In fact, structurally equivalent optimization terms are evaluated to isomorphic \nhg s (soundness). 

For completeness, given a \nhg\ $\ngraph{\eta}{G}$, we can construct an equivalent term in normal form (regarded as canonical representative of its $\tequiv$-class, by \cref{rem:can-rep}) as follows. We encode each edge $e$ of $G$ as an atomic term with label $lab_G(e)$. The arguments of this term are the names of the interface vertices of $e$, and arbitrary ones for non-interface vertices. Finally, we form the parallel composition of all these terms and we restrict the names that are not in the interface. Every choice of restricted names is valid: $\alpha$-conversion guarantees that all possible encodings of $\ngraph{\eta}{G}$ are in the same structural congruence class.

\subsection{Tree decompositions}

In graph theory, we have the well-known notion of \emph{tree decomposition} of a graph \cite{RobertsonS84}, which can be understood as a way of parsing a graph. We report the definition by \cite{Kloks94}.
\begin{definition}[Tree decomposition]
\label{def:td}
A \emph{tree decomposition} (TD) of a hypergraph $G$ is a pair $(T,X)$, where $T = (N,A)$ is a tree (i.e., an undirected, acyclic graph) and $X = \{X_n\}_{n \in N}$ is a family of subsets $X_n \subseteq V_G$, one for each node of $N$, such that: (a) $\bigcup_{n \in N} X_n = V_G$; (b) for all $e \in E_G$ there is $n \in N$ such that $a_G(E) \subseteq X_n$; (c) for all $v \in V_G$, the set of nodes $\{ n \in N \mid v \in X_n \}$ induces a subtree of $T$.
\end{definition}
In our framework, ordinary hypergraphs (without isolated vertices) can naturally be seen as \nhg s with empty interface, and each of their TDs can be represented as an optimization term.
\begin{theorem}
Every TD for $G$ induces an optimization term $t$ such that $\deng{t} \cong \ngraph{\uparrow}{G}$, where $\uparrow \colon V_G \to \names$ is the nowhere defined function.
\label{thm:td-term}
\end{theorem}
The idea is constructing $t$ via a visit of $T$ from a chosen root $r$. We first associate the induced subgraph $G_n$ of $G$ to $X_n$, for each node $n \in N$. Each time a new node $n$ is expanded in the visit, we generate the following subterms of $t$, all in parallel: one term representing edges and nodes of $G_n$ not already in $t$; the subterms corresponding to $n$'s children. Correctness of the translation is guaranteed by (a)-(c) of \cref{def:td}.
Notice that, since all variables are restricted, choosing a different root amounts to rearranging restrictions. By soundness, this operation results in terms with isomorphic images via $\deng{-}$, all isomorphic to $G$.

\section{Representing and solving optimization problems}
\label{sec:costfun-alg}

We now show how typical optimization problems can be represented and solved in our algebraic framework.\footnote{A different, more efficient, setting will be described in \cref{ssec:parking}.}
Suppose we have $n$ atomic problems $A_1,\dots,A_n$ whose variables can be assigned values in $\Vals$, and we want to minimize a function of the form
\[
	\sum_{1 \leq i \leq n} c_{A_i}(\tilde{x}_i)
\]
where each $c_{A_i}(\tilde{x}_i) \colon \Vals^{|\tilde{x}_i|} \to \RealInf$, for $i=1,\dots,n$, gives a cost to each variable assignment for the problem $A_i$; an infinite cost represents a forbidden assignment.

The problem can be represented as the following term in normal form
\[
	p = (\tilde{x})( A_1(\tilde{x}_1) \parallel \dots \parallel A_n(\tilde{x}_n) ) \qquad \text{where} \quad \tilde{x} = \tilde{x}_1 \cup \dots \cup \tilde{x}_n
\]
and the computation of the optimal cost as a function
\[
    \denc{p} \colon (\names \to \Vals) \to \RealInf
\]
giving a cost to each assignment of variables. More precisely, its computation is performed by assigning values to the free variables of $p$ (discarding assignments to other variables), and minimizing w.r.t.\ bound ones. Typically we have $fn(p) = \varnothing$, so minimization is performed w.r.t.\ all variables.

Formally, we take an algebra for the optimization specification formed by \emph{cost functions} $\phi \colon (\names \to \Vals) \to \RealInf$, where we interpret optimization terms. For any assignment of variables $\rho \colon \names \to \Vals$, the interpretation of constants is
\[
    \denc{A_i(x_1,\dots,x_n)}\rho = c_{A_i}(\rho(x_1),\dots,\rho(x_n)) \qquad\qquad
    \denc{nil}\rho = 0
\]
and complex terms are recursively interpreted as follows
\[
		\denc{p_1 \parallel p_2} \rho = \denc{p_1} \rho + \denc{p_2} \rho
	\qquad
	    \denc{ (x) p } \rho = \min_{v \in \Vals} \denc{p}(\rho[x \mapsto v])
		\qquad
		\denc{p \pi} \rho = \denc{p} (\rho \circ \pi^{-1} )
\]
We have the following property, which comes from the theory of permutation algebras.
\begin{property}
$supp(\denc{p}) \subseteq fn(p)$.
\label{prop:den-supp}
\end{property}

We introduce a condition on cost functions, called \emph{compactness}. A compact $\phi$ depends only on a ``few'' variables. This is essential to compute and store cost functions in an efficient way, as we will see later.
\begin{property}[Compactness]
We say that  $\phi \colon (\names \to \Vals) \to \RealInf$ is \emph{compact} if $\funres{\rho}{supp(\phi)} = \funres{\rho'}{supp(\phi)}$ implies $\phi \rho = \phi \rho'$, for all $\rho,\rho' \colon \names \to \Vals$.
\label{prop:supp}
\end{property}
When considering a term $p$, by \cref{prop:supp}, $\denc{p}$ is compact if it depends only on assignments to free variables of $p$. This is clearly the case for the interpretation of constants, and can be shown by structural induction for complex terms. Notice that this property is not true for the whole algebra of functions $\phi \colon (\names \to \Vals) \to \RealInf$, but only for the subalgebra in the image of $\denc{-}$.

Canonical forms and normal forms of a term always have the same cost function. This is thanks to the following proposition, which is a direct consequence of cost functions forming an algebra of the optimization specification.
\begin{proposition}
If $p \equiv q$ then $\denc{p} = \denc{q}$.
\end{proposition}

\subsection{Computational complexity of cost functions}
\label{ssec:complexity}

Although structurally congruent terms have the same cost functions, these functions may be computed in different ways, each possibly with a different computational cost. In fact, the position of restrictions inside a term determines a strategy for variable elimination. As already mentioned, finding the best one amounts to giving a solution for the secondary optimization problem.

We introduce a notion of complexity for an optimization problem $p$, similar to the one of \cite{BerteleB73}, estimating the cost of computing its value $\denc{p}$. This is given by the function with greatest ``size'' encountered while inductively constructing $\denc{p}$, the size being given by the variables in the support, that are the only ones determining the value of $\denc{p}$ (\cref{prop:supp}).

Formally, the complexity of $p$, written $\cost{p}$, is recursively defined as follows
\[
    \cost{A(\tilde{x})} = |\tilde{x}|  
    \qquad
    \cost{nil} = 0
    \qquad
     \cost{(x)p} = \cost{p} 
     \qquad
    \cost{p \parallel q} = \max \, \{ \cost{p} , \cost{q}, |fn(p \parallel q )| \}
\]
The interesting cases are $(x)p$ and $p \parallel q $: the computation of $\denc{(x)p}$ relies on that of $\denc{p}$, whose support may be bigger, so we set the complexity of $(x)p$ to that of $p$; computing $\denc{p \parallel q}$ requires computing $\denc{p}$ and $\denc{q}$, but the support of the resulting function is the union of those of $p$ and $q$, so we have to find the maximum value among $\cost{p}$, $\cost{q}$ and the overall number of free variables.

Complexity is well-defined only for hierarchical terms: applying \axse\ to choose a different variable elimination strategy may change the complexity. Consider, for instance, the following term in normal form
\[
   p = (x_1)(x_2)(x_3)(A(x_1,x_2) \parallel B(x_2,x_3)) \enspace ;
\]
we have $\cost{p} = 3$, but if we take a canonical form
\[
    q = (x_2)((x_1)A(x_1,x_2) \parallel (x_3)B(x_2,x_3))
\]
we have $\cost{q} = 2$. Indeed, we have the following results for hierarchical terms.
\begin{lemma}
Given $(x)( p \parallel q)$, with $x \notin fn(q)$, we have
$\cost{(x)p \parallel q} \leq \cost{(x)(p \parallel q)}$. 
\label{lem:axse-complexity}
\end{lemma}
As an immediate consequence, all the canonical forms of a term always have lower or equal complexity than the normal form.
\begin{theorem}
Given a term $p$, let $n$ be its normal form. Then, for all canonical forms $c$ of $p$ we have $\cost{c} \leq \cost{n}$.
\label{th:can-norm-cost}
\end{theorem}

\subsection{Example}
\label{ssec:ex}
Consider two problems $A$ and $B$, with two variables each, ranging over $\{d_1,d_2\}$. Their cost functions are shown in Tables \ref{tab:ex-A} and \ref{tab:ex-B}. We consider the optimization problem that consists in finding the minimal value of $A(x_1,x_2) + B(x_2,x_3)$. 

\begin{table}[t]
\centering
\begin{subtable}[t]{.3\linewidth}
\caption{$\denc{A(x_1,x_2)}$}
\label{tab:ex-A}
\centering
\begin{tabular}{ccc}
$x_1$ & $x_2$ & $cost$
\\
\hline\hline
$d_1$ & $d_1$ & 7 \\
$d_1$ & $d_2$ & 5 \\
$d_2$ & $d_1$ & $\infty$ \\
$d_2$ & $d_2$ & 2 \\
\end{tabular}
\end{subtable}
\qquad
\begin{subtable}[t]{.2\linewidth}
\caption{$\denc{B(x_2,x_3)}$}
\label{tab:ex-B}
\centering    
\begin{tabular}{ccc}
$x_2$ & $x_3$ & $cost$
\\
\hline\hline
$d_1$ & $d_1$ & 9 \\
$d_1$ & $d_2$ & 1 \\
$d_2$ & $d_1$ & 6 \\
$d_2$ & $d_2$ & 13 \\
\end{tabular}
\end{subtable}
\qquad
\begin{subtable}[t]{.3\linewidth}
\caption{$\denc{(x_1)A(x_1,x_2)}$}
\label{tab:ex-(x1)A}
\centering
\begin{tabular}{cc}
$x_2$ & $cost$
\\
\hline\hline
$d_1$ & $\min \{7,\infty\} = 7$ \\
$d_2$ & $\min \{5,2\} = 2$
\end{tabular}
\end{subtable}

\bigskip
\begin{subtable}[t]{.4\linewidth}
\caption{$\denc{(x_3)B(x_2,x_3)}$}
\label{tab:ex-(x3)B}
\centering    
\begin{tabular}{cc}
$x_2$ & $cost$
\\
\hline\hline
$d_1$ & $\min \{9,1\} = 1$ \\
$d_2$ & $\min \{6,13\} = 6$
\end{tabular}
\end{subtable}
\qquad
\begin{subtable}[t]{.4\linewidth}
\caption{$\denc{(x_1)A(x_1,x_2) \parallel (x_3)B(x_2,x_3)}$}
\label{tab:ex-par}
\centering
\begin{tabular}{cc}
$x_2$ & $cost$
\\
\hline\hline
$d_1$ & $7 + 1 = 8$ \\
$d_2$ & $2 + 6 = 8$
\end{tabular}
\end{subtable}

\caption{Cost functions for the problems in the example.}
\label{tab:ex}
\end{table}

As we already saw, the term in canonical form representing the problem is
\[
	p = (x_2) ( (x_1) A(x_1,x_2) \parallel (x_3) B(x_2,x_3) )
\]
We now show how $\denc{p}$ can be computed. We proceed in a bottom-up order, from atomic subterms to increasingly complex terms. This is close to a dynamic programming algorithm, as it allows computing and storing a (finite, thanks to the compactness property) representation of cost functions once and for all. \Cref{tab:ex} show such finite representations in a tabular form.
We perform the following optimization steps, each corresponding to an operator of the syntax:

\begin{enumerate}
\item $\denc{(x_1)A(x_1,x_2)}$ and $\denc{(x_3)B(x_2,x_3)}$ are computed by minimizing $\denc{A(x_1,x_2)}$ and $\denc{B(x_2,x_3)}$ w.r.t.\ $x_1$ and $x_3$ respectively (Tables \ref{tab:ex-(x1)A} and \ref{tab:ex-(x3)B}). Notice that these functions can be computed in parallel.
\item $\denc{(x_1)A(x_1,x_2) \parallel (x_3)B(x_2,x_3)}$ is computed by evaluating $\denc{(x_1)A(x_1,x_2)}$ and $\denc{(x_3)B(x_2,x_3)}$ on the same value for $x_2$, and then summing up the results (\Cref{tab:ex-par}).
\item Finally, $\denc{(x_2)( (x_1)A(x_1,x_2) \parallel (x_3)B(x_2,x_3))}$ is computed by minimizing the function of step 2) w.r.t.\ $x_2$.
\end{enumerate}
The last step gives the overall minimal value 8. By looking at Tables in a top-down order, from \ref{tab:ex-par} to \ref{tab:ex-A}, each time picking those variable assignments that contributed to the cost, one can recover the corresponding optimal assignment(s) for $x_1,x_2$ and $x_3$, namely $d_1,d_1,d_2$ and $d_2,d_2,d_1$. 
\section{Dynamic programming on hierarchical \nhg s}

The existence of an algebra of \nhg s allows us to recursively compute cost functions for these graphs: given a \nhg\ $\ngraph{\eta}{G}$ and an optimization term such that  $\deng{p} = \ngraph{\eta}{G}$, we can compute its cost function in the style of \cref{sec:costfun-alg}.

However, the information about the variable elimination strategy cannot be recovered from the \nhg\ itself.
In order to do this, we need to introduce the graphical counterpart of hierarchical terms, which we call \emph{hierarchical} \nhg s. They are trees that describe the structure of a \nhg\ $\ngraph{\eta}{G}$ in terms of nested components. These trees are such that:
\begin{itemize}    
    \item the root is the discrete hypergraph formed by the interface vertices of $G$;
    \item each internal node $n$ is a discrete subgraph of $G$;
    \item leaves are hypergraphs with a single hyperedge of $G$;
    \item there is an arc from $G$ to $G'$ whenever $G \subset G'$.
\end{itemize}
The intuition is that each internal node $n$ of the tree is a component of $\ngraph{\eta}{G}$ that exposes some additional vertices and includes all the components in the subtree rooted in $n$. Leaves are basic components, i.e., hyperedges.

\begin{figure}[t]
\begin{tabular}{m{.5\textwidth}m{.4\textwidth}}
\[
    (x_2) ( (x_1) A(x_1,x_2) \parallel (x_3) B(x_2,x_3) )
\]
&
\includegraphics[scale=.35]{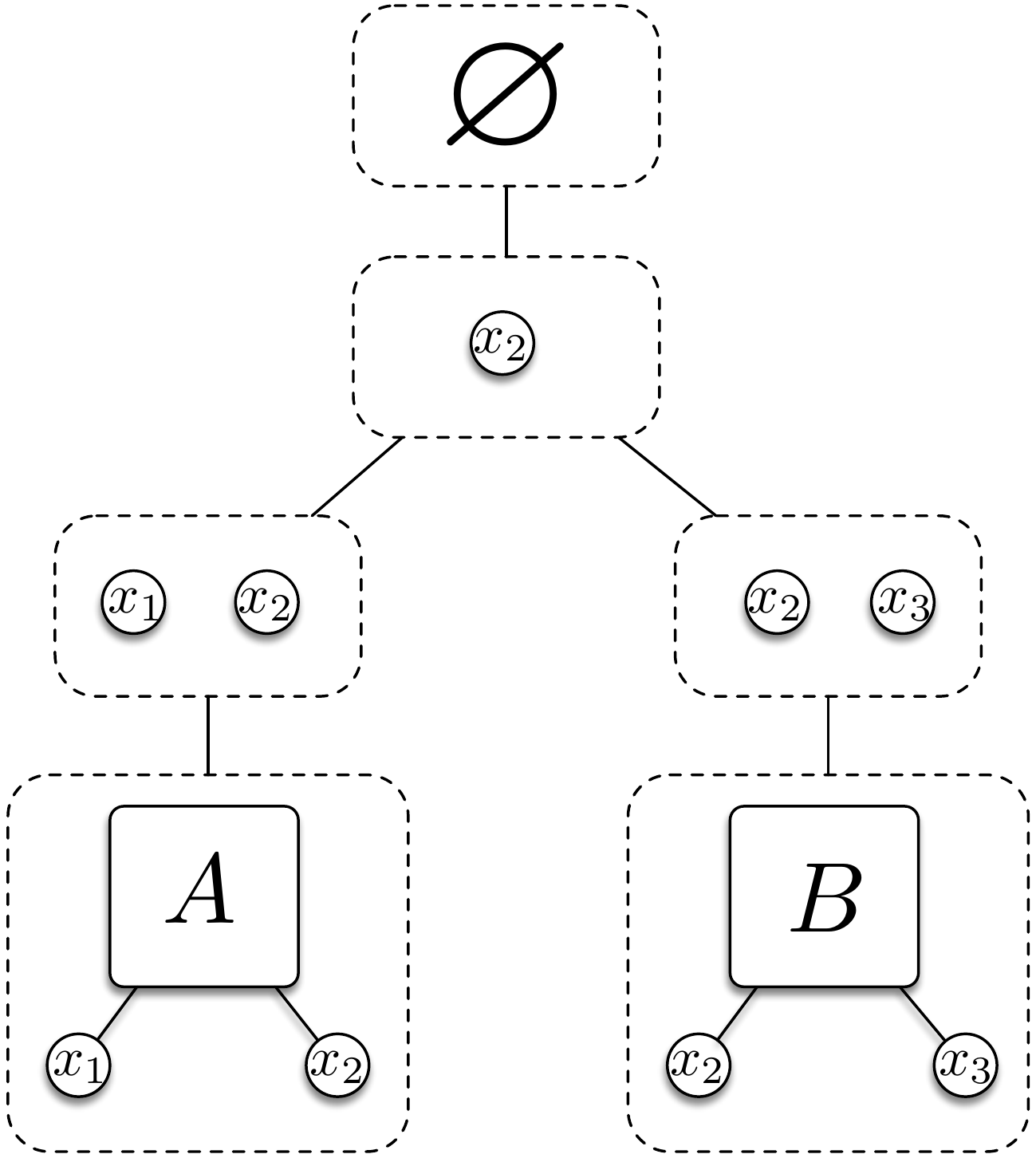}
\end{tabular}
\caption{A hierarchical term and the corresponding hierarchical \nhg.}
\label{fig:hier-corr}
\end{figure}

The correspondence between hierarchical terms and hierarchical \nhg\ graphs is exemplified in \cref{fig:hier-corr}. The scope of each restriction determines a component in the tree, where a vertex for the restricted name is added. For convenience, we used the same name for restricted variables and corresponding non-interface vertices, but the latter, as in ordinary \nhg s, are actually up to $\alpha$-conversion. A top-down visit of the tree amounts to ``opening'' scopes and revealing their names. 

As hierarchical terms, hierarchical \nhg s describe a solution for the secondary optimization problem. It is possible to show that hierarchical \nhg s form an algebra for the hierarchical optimization specification. As seen in the example, each hierarchical term corresponds to a hierarchical \nhg. We also have the opposite correspondence. 

\begin{proposition}
Each hierarchical \nhg\ can be represented as a hierarchical term.
\label{prop:hier-nhg-term}
\end{proposition}
Exploiting this correspondence, we can interpret hierarchical \nhg s as evaluations of cost functions with a specific variable elimination strategy, which can be implemented via dynamic programming.
\begin{remark}
In \cref{rem:span} we have characterized \nhg s as spans of hypergraph homomorphisms. Interestingly, exploiting this characterization we can recover the \nhg\ of which a given hierarchical \nhg\ is the decomposition. In fact, a hierarchical \nhg\ can be regarded as a span where the right part is a diagram $T$ made of a ``tree'' of graph embeddings.
\[
    \xymatrix{
        [\names] & [img(\eta)] \ar@{_{(}->}[l]_-{\eta_l} \ar@{^{(}->}[r]^-{\eta_r} & T
    }
\]
Here $\eta_r$ maps interface vertices to themselves in the root of $T$. Notice that the morphisms of $T$ tell which are the interface vertices in the nodes of $T$. In order to ``paste'' together leaf hypergraphs of $T$, we can make a colimit of $T$ in the category of graphs and their morphisms. The result is the disjoint union of such hypergraphs, where vertices that are images of the same interface one, along the morphisms of $T$, are identified. 
\end{remark}

\subsection{The parking optimization problem}
\label{ssec:parking}

We now introduce the parking optimization problem and we apply our approach to it. The parking optimization problem consists in finding the best parking zone for each vehicle of an ensemble, that is a group of vehicles with similar features. This can be formalized as follows. Assume a set of parking zones $P = \{ A,B,\dots \}$ and of car variables $C = \{ x,y,\dots \}$, and two functions:
\begin{itemize}
	\item $c \colon P \to \Nat$, assigning a \emph{capacity} to every zone;
	\item $F \colon C \to P \to \RealInf$, specifying the cost $F(x)(A)$ for $x$ to park in $A$.
\end{itemize}
Given an assignment $\rho \colon C \to P$ of cars to zones, let $\rho_A = \{ x \mid \rho (x) = A \}$. We want to find an assignment $\rho$ such that $|\rho_A| \leq c(A)$, for all $A \in P$, minimizing
\[
	\sum_{x \in \names} F (x)(\rho(x))
\]

The problem can be specified in the style of \cref{sec:desc}. Here a term $p$ represents a parking system: $A(x_1,\dots,x_n)$ means that $x_i$ might be parked in $A$; $(x)p$ means that car $x$ cannot be parked outside of $p$, so it must have a parking spot in one of the zones of $p$. In general, a term $p$ represents a part of the system made of one or more parking zones.

In \cref{sec:costfun-alg} we presented an algebra of cost functions for typical optimization problems. Here we introduce another, ``more efficient'' algebra which, nevertheless, fits into our algebraic framework and can be used to evaluate optimization terms. To each parking system $p$ we associate a function
\[
    \denc{p}: \Parts(fn(p)) \to \RealInf
\] 
The intended meaning of $\denc{p}X$ is the cost of parking in $p$ cars $X \subseteq fn(p)$ and all cars corresponding to variables restricted in $p$. Notice that the evaluation function $\denc{-}$ is quite different from  that of \cref{sec:costfun-alg}. Here assignments do not fix the values of variables, i.e., the parking zones where cars are allocated, but only their positions with respect to the present $p$. However, our framework is able to accommodate also this more efficient setting.

To avoid handling polymorphic functions, we automatically extend functions $\denc{p}$ to the whole $\names$, namely $\denc{p} \colon \Parts(\names) \to \RealInf$, by letting 
\begin{equation}
    \denc{p}X = \denc{p}(X \cap fn(p)) . \label{eq:var-supp}
\end{equation}
However, this function is still determined by subsets of $fn(p)$, so it admits a finite representation that can be efficiently computed and stored. Formally, properties \ref{prop:den-supp} and \ref{prop:supp} hold for extensions to $\names$ (regarded as functions $(\names \to \{0,1\}) \to \RealInf$).

Cost functions can be defined by recursion on the structure of systems. As mentioned, it is enough to define their action on a subset $X$ of their support. We have
\[
	\denc{ A(\tilde{x}) }X = 
	\begin{cases}
		\sum_{x \in X} F(x)(A) & |X| \leq c(A) \\
		\infty & \text{otherwise}
	\end{cases}
\]
meaning that cars $X \subseteq \tilde{x}$ can be parked in zone $A$ iff their number is at most the capacity of $A$. For $nil$ we simply have $\denc{nil}X = 0$.

Then we have
\[
 \denc{p \parallel q} X = \min_{\{X_1,X_2\} \in \Parts_2(X)} \left\{ \denc{p}X_1 + \denc{q}X_2 \left| 
\begin{array}{l}
X_1 \subseteq fn(p), \\
X_2 \subseteq fn(q)
\end{array}
\right.
\right\}    
\]
where $\Parts_2(X)$ are the partitions in two sets of $X \subseteq fn(p \parallel q)$. Here, to park cars $X$ in component $p \parallel q$, one has to park each of them in either component $p$ or component $q$, but not in both. Thus the best option must be chosen. Finally, we have
\[
\denc{(x)p}X = \denc{p}(X \cup \{x\}) .
\]
Here it is required that car $x$ is parked in component $p$.

Typically, the whole system $s$ has no free names. Thus $\llbracket s \rrbracket \varnothing$ is a real number, the total minimized cost, or $\infty$ if the problem has no solution.

In order to have a proper theory of cost functions, we have to show that we have indeed defined a model of the optimization specification. Let $X\pi$ be the element-wise application of a permutation $\pi \colon \names \to \names$ to $X \subseteq \names$. Then we have the following theorem.

\begin{theorem}
Cost functions $\phi \colon \Parts(\names) \to \RealInf$ satisfying (\ref{eq:var-supp}) form a model of the optimization specification, together with the given interpretation of operators and the permutation action $(\phi \pi)X = \phi (X \pi^{-1})$.
\label{th:bool-model}
\end{theorem}

\subsection{Dynamic programming algorithm}

Consider the scenario with three possible parking zones $A, B, C$ and three cars $x_1, x_2$ and $x_3$. We assume the following values for $F$ and $c$.
\begin{alignat*}{5}
    F(x_1)(A) &= 3 && \qquad & F(x_1)(B) &= \infty && \qquad & F(x_1)(C) &= \infty  \qquad c(A) = 2 \qquad c(B) = 2 \qquad c(C) = 2\\
    F(x_2)(A) &= 4 &&& F(x_2)(B) &= 6 &&& F(x_2)(C) &= \infty \\
    F(x_3)(A) &= \infty &&& F(x_3)(B) &= 4 &&& F(x_3)(C) &= 1 
\end{alignat*}

\begin{figure}[t]
\begin{tabular}{m{.46\textwidth}m{.48\textwidth}}
\begin{center} {\bf Normal} \end{center}
&
\begin{center} {\bf Canonical} \end{center}
\\
$ p = (x_1)(x_2)(x_3) (A(x_1,x_2) \parallel B (x_2,x_3) \parallel C(x_3)) $
&
$
c = (x_2)((x_1)A(x_1,x_2) \parallel (x_3) (B(x_2,x_3) \parallel C(x_3)))
$
\\[3ex]
\centering
\includegraphics[scale=.35]{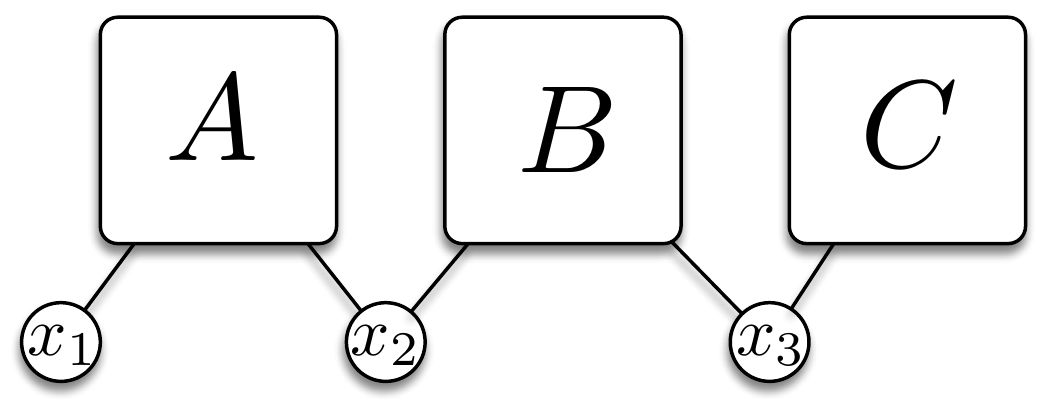}
&
\centering
\includegraphics[scale=.35]{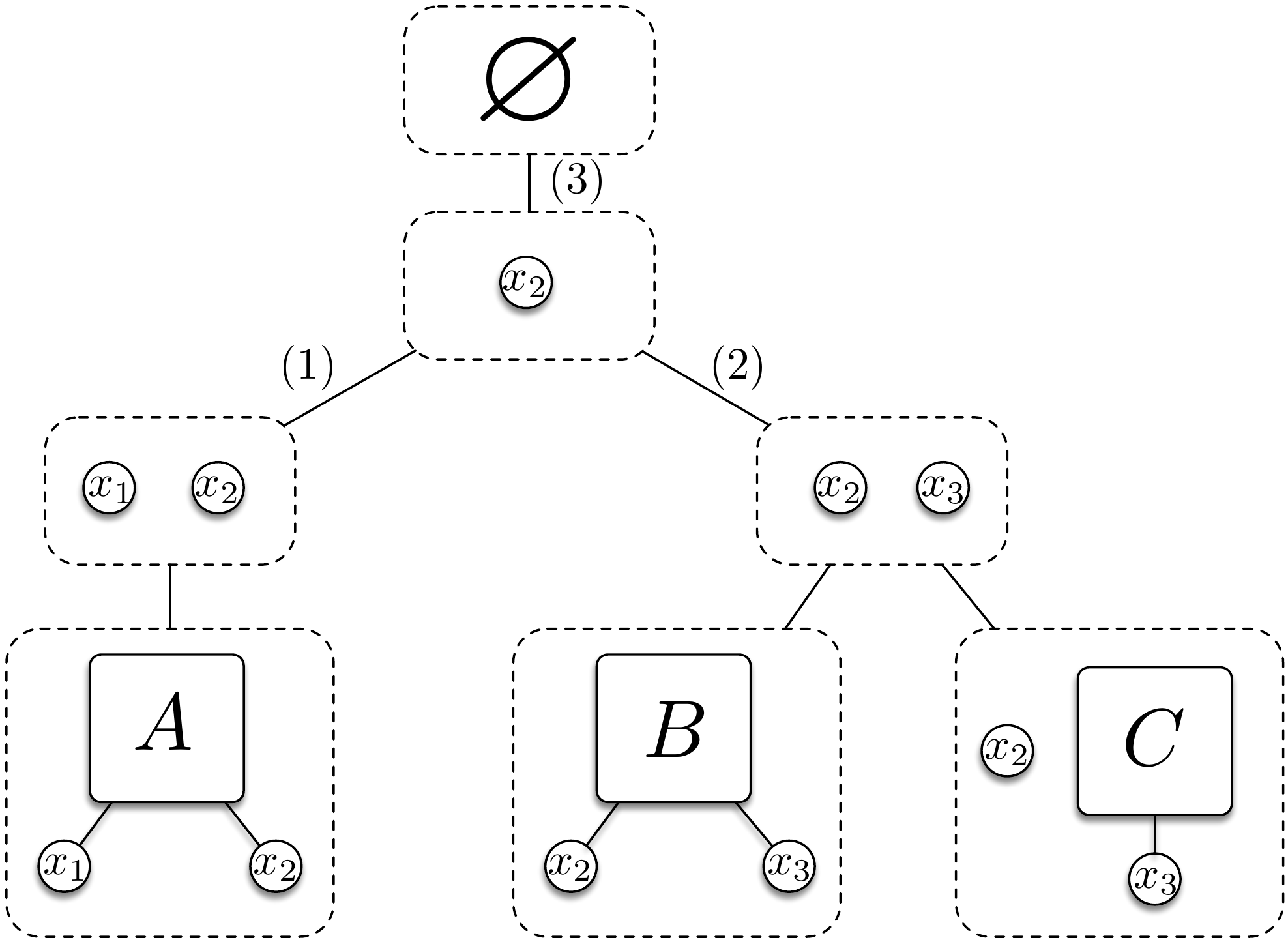}
\end{tabular}
\caption{Graphical and corresponding term representation of a parking problem.}
\label{fig:park}
\end{figure}
In \cref{fig:park}, on the left side, we show the term in normal form, and the corresponding \nhg, modeling the system. We want to compute the cost function $\denc{p}$ using dynamic programming. The crucial property is compactness of cost functions: $\denc{p}$ can be represented as a table of size $|fn(p)|$\footnote{The actual size is $\mathcal{O}(2^{|fn(p)|})$, but we show the exponent, as $2^x \leq 2^y$ iff $x \leq y$.}. Although a problem is typically specified as a term in normal form, we consider its canonical form $c$, shown on the right side of \cref{fig:park}, because its complexity is equal or lower (\cref{th:can-norm-cost}). We assume that all occurrences of $nil$ have been eliminated via structural congruence.
\begin{table}[t]
\caption{Example tables. Parameters of atomic subterms are often omitted.}
\label{tab}
\centering
\begin{subtable}[t]{.3\linewidth}
\caption{$\denc{A(x_1,x_2)}$}
\label{tab:TA}
\centering
\begin{tabular}{cccc}
$x_1$ & $x_2$ & $cost $
\\
\hline\hline
$\checkmark$ & $\checkmark$ & 7 \\
$\checkmark$ & $-$ & 3 \\
$-$ & $\checkmark$ & 4 \\
$-$ & $-$ & 0
\end{tabular}
\end{subtable}
\begin{subtable}[t]{.3\linewidth}
\caption{$\denc{B(x_2,x_3)}$}
\label{tab:TB}
\centering
\begin{tabular}{cccc}
$x_2$ & $x_3$ & $cost$ 
\\
\hline\hline
$\checkmark$ & $\checkmark$ & 10 \\
$\checkmark$ & $-$ & 4 \\
$-$ & $\checkmark$ & 6 \\
$-$ & $-$ & 0
\end{tabular}
\end{subtable}
\begin{subtable}[t]{.3\linewidth}
\caption{$\denc{C(x_3)}$}
\label{tab:TC}
\centering
\begin{tabular}{cc}
$x_3$ & $cost$ 
\\
\hline\hline
$\checkmark$ & 1\\
$-$ & 0
\end{tabular}
\end{subtable}

\bigskip

\begin{subtable}[t]{.3\linewidth}
\caption{$\denc{(x_1)A(x_1,x_2)}$}
\label{tab:Tx1A}
\centering
\begin{tabular}{cc}
$x_2$ & $cost$ 
\\
\hline\hline
$\checkmark$ & 7
\\
$-$ & 3
\end{tabular}
\end{subtable}
%
%
\begin{subtable}[t]{.6\linewidth}
\caption{$\denc{(x_3) ( \, B(x_2,x_3) \parallel C(x_3) \, )}$}
\label{tab:Tx2}
\centering
\begin{tabular}{c|cc|c|cc}
\multicolumn{1}{c}{$x_2$} & \multicolumn{2}{c}{$\denc{B}$} & \multicolumn{1}{c}{$\denc{C}$} & $\denc{B} + \denc{C}$ & $cost$
\\
\hline\hline
\multirow{3}{*}[-2ex]{$\checkmark$} & $x_3$ & $x_2$ & $x_3$ & & \multirow{3}{*}[-2ex]{7}
\\
\cline{2-4}
& $\checkmark$ & $\checkmark$ & $-$ & 10 \\
& $-$ & $\checkmark$ & $\checkmark$ & 7 \\
\hline
\multirow{2}{*}{$-$} & $\checkmark$ & $-$ & $-$ & 4 & \multirow{2}{*}{1}
\\
& $-$ & $-$ & $\checkmark$ & 1
\end{tabular}
\end{subtable}

\vspace{4ex}
\begin{subtable}[t]{\linewidth}
\caption{$\denc{(x_2)( \ (x_1)A(x_1,x_2) \parallel (x_3) ( \, B(x_2,x_3) \parallel C(x_3) \, ) \ )}$}
\label{tab:Tx3}
\centering
\begin{tabular}{|c|c|cc}
\multicolumn{1}{c}{$\denc{(x_1)A}$}  & \multicolumn{1}{c}{$\denc{(x_3)(B \parallel C)}$} & $\denc{(x_1)A} +  \denc{(x_3)(B \parallel C)}$ & $cost$
\\
\hline
\hline
$x_2$ & $x_2$ && \multirow{3}{*}[-2ex]{8}
\\
\cline{1-2}
$\checkmark$ & $-$ & 8 \\
$-$ & $\checkmark$ & 10
\end{tabular}
\end{subtable}
\end{table}

We propose a dynamic programming algorithm that is driven by the hierarchical \nhg\ that $c$ describes, shown in the right side of \cref{fig:park}. This algorithm operates as the one in \cref{ssec:ex}, but tables are computed using the different interpretation of operators on cost functions, introduced in \cref{ssec:parking}.

The algorithm starts from the cost functions for the leaves. These are shown in \Crefrange{tab:TA}{tab:TC}, where the leftmost columns indicates whether a car is parked inside ($\checkmark$) or outside ($-$) each zone. They are computed as described in \cref{ssec:parking}: e.g., for $\denc{A(x_1,x_2)}$, the cost for each row is $\denc{A(x_1,x_2)}X$, where $X$ are the variables marked with $\checkmark$ in that row. Then, the algorithm performs a bottom-up visit of the tree and eliminates variables accordingly. More precisely, whenever an edge from $G$ to $G'$ is traversed, with $G$ and $G'$ discrete hypergraphs, variables $G \setminus G'$ are eliminated. In the following we show the elimination steps, also indicated in \cref{fig:park}: 
\begin{enumerate}[(1)]
\item \emph{Elimination of $x_1$}: Table $\denc{(x_1)A(x_1,x_2)}$ (\ref{tab:Tx1A}), with only one column $x_2$, is computed by forcing $x_1$ to be inside $A$;
\item \emph{Elimination of $x_3$}: the table $\denc{(x_3) (B(x_2,x_3) \parallel C(x_3))}$
is computed: \Cref{tab:Tx2} shows values for $x_2$, the partitions considered when computing the output cost, and the final cost. Notice that this and the previous step could be executed in parallel. This fact comes immediately from terms $(x_1)A(x_1,x_2)$ and $(x_3)(B(x_2,x_3) \parallel C(x_3))$ being composed in parallel in $(x_1)A(x_1,x_2) \parallel (x_3)(B(x_2,x_3)\parallel C(x_3))$.
\item \emph{Elimination of $x_2$}: finally, the Table $\denc{p}$ (\ref{tab:Tx3}) is computed, by comparing costs of parking $x_2$ inside either $(x_1)A(x_1,x_2)$ or $(x_3)( \, B(x_2,x_3) \parallel C(x_3) \, )$.
\item \emph{Optimal variable assignment:} tracking back through the Tables we find:
\begin{itemize} 
    \item $x_2$ inside $(x_1)A(x_1,x_2) \parallel (x_3)(B(x_2,x_3) \parallel C(x_3))$;
    \item $x_2$ inside  $(x_1)A(x_1,x_2)$;
    \item $x_2$ inside  $A(x_1,x_2)$ with cost 4;
    \item $x_1$ inside $A(x_1,x_2)$ with cost 3;
    \item $x_3$ inside  $B(x_2,x_3) \parallel C(x_3)$;
    \item $x_3$ inside  $C(x_3)$ with cost 1.
\end{itemize}
\end{enumerate}
Notice that, in general, the outcome of the algorithm may be $\infty$, whenever there is no car assignment to parking zones that respect capacities.

\section{Conclusion}

In the paper we have introduced two process algebra-like specifications for the description of optimization problems. The more abstract version (which includes the scope extension axiom) defines (hyper) graphs, where vertices are variables, and edges are tasks whose costs depend on the values of the adjacent variables. Dropping the above axiom yields a specification corresponding to different parsing trees of the given graph. Choosing a particular tree corresponds to selecting a dynamic programming strategy for the given problem, whose execution can be carried on via a bottom up visit of the tree. We apply our approach to the parking optimization problem developed, in collaboration with Volkswagen, in the ASCENS e-mobility case study.

The idea of exploiting graphs to decompose and solve various kinds of problems is not new. In \cite{CourcelleM93} graphs are represented as elements of an algebra and monadic second-order properties are evaluated on them. In \cite{BodlaenderK08} dynamic programming algorithms are derived from (\emph{nice}) tree decompositions of graphs. In \cite{BlumeBFK13} tree decompositions are represented as in a category of spans and cospans, and then as terms of an algebraic specification. Our approach has the following advantages w.r.t.\ the cited ones: 
\begin{itemize}
    \item Our algebraic specification is simpler, but nonetheless expressive. In fact, variable elimination strategies can be represented via restrictions. Moreover, we have a graphical representation of such strategies as hierarchical \nhg s, which can be regarded as very simple tree decompositions.
    \item Our algebras are permutation algebras, which provide: (a) a state-of-the-art treatment of $\alpha$-conversion and of freshness requirements; (b) a uniform and general definition of domain given by the notion of support. Operations are defined on the whole set of names, so they are independent of the actual interface (support), unlike \cite{BlumeBFK13}. Moreover, the notion of support automatically defines the sizes of the tables employed in the dynamic programming implementation. 
\end{itemize}

The following lines of research are also related to our work. 
Bistarelli, Montanari and Rossi deal with SCSP \cite{BistarelliMR97} and its combination with logic programming \cite{BistarelliR01,BistarelliMR02} and concurrency \cite{BistarelliMR06}. They give an interpretation of constraints over certain semirings, e.g., the tropical semiring, as we do here. However, operations on constraints are defined point-wise using the semiring operations: the approach is too restrictive, e.g., it does not easily accommodate the case study shown in this paper. A direct connection between (logical) CSP and dynamic programming is shown in \cite{MontanariR91}.
Dechter in \cite{Dechter99} introduces bucket elimination as a
general solution technique for a variety of problems: it consists in a strategy of problem reduction employing a convenient elimination ordering of variables and constraints. The associated technique of conditioning search allows for approximated versions of the bucket elimination approach.
Kohlas and Pouly in \cite{KohlasP11} suggest \emph{valuation algebras} as a foundation for a general view of information processing. They define axioms for valuation algebras, consider a number of instantiations and provide generic inference algorithms for their processing. Our approach is similar, but more direct, being based on a simple process algebra specification and on a bottom up visit of a tree of graphs satisfying the specification. 
In \cite{CorradiniMR94} distributed systems are represented as \emph{CHARMs} (Concurrency and Hiding in an Abstract Rewriting Machine), that are hypergraphs with a global and a local part. They form an algebra including edges and vertices restriction. Our algebras and \nhg s are similar, but we do not need edges restriction.

Dynamic programming evaluations are particularly interesting for autonomic systems, as studied by the ASCENS project, where the actual behavior often consists, typically for the dynamic programming case, of propagating local knowledge to obtain global knowledge and getting it back for local decisions.  When dealing with global problems, however, the complexity of the dynamic programming algorithms can grow exponentially even for graphs of limited complexity. Consider a rectangular grid of size $n$, with vertices labeled by variables, and edges by cost functions with two arguments. It is shown in \cite{MartelliM72} that its complexity is exponential in $n$. There are efficient algorithms for finding the optimal elimination order of vertices in a graph, but they deal with specific cases (e.g., Gaussian elimination \cite{Dahlhaus02,Yannakakis81}).
Thus approximation techniques are quite relevant, in particular when a good global solution, possibly not optimal, is still acceptable. 

Several heuristic techniques can be experimented. For instance, for the parking problem we could restrict the number of possible zones for each car, taking the best $k$ of them. Then if the optimal solution would include a choice worse than $k$ for some car, the solution found, if any, would not be optimal. However, at least no client would be treated too badly. Another, quite general, approximation technique would be to artificially reduce the dimensions of tables by decomposing high dimensional ones into the sums of a few lower dimensional tables. The latter can be computed minimizing the mean square error \cite{Montanari71}. The storage reduction can be propagated in such a way to reduce substantially the overall complexity.

An interesting piece of future work would be to extend our approach to graphs which are incrementally modified, e.g., extended, at run time. The resulting scenario could consist of a (soft) (concurrent) constraint component together with a mobile pi-calculus-like process algebra component. A good example of this combination is cc-pi \cite{BuscemiM07}. Other aspects should be investigated in a precise way: the correspondence between classes of terms (normal,canonical) and \nhg s; the nominal structure of \nhg s.

\bibliographystyle{eptcs}
\bibliography{GAM15}

\end{document}